\documentclass{ptephy_v1}

\preprintnumber{XXXX-XXXX} 

\usepackage{graphicx}
\usepackage{epstopdf}
\usepackage{amsmath}
\usepackage{amsfonts}
\usepackage{amssymb}
\usepackage{bm}

\begin{document}
\title{ Pion--nucleon sigma term $\sigma_{\pi N}$ and deeply bound pionic atoms}
\date{\today}

\author[1,*]{Natsumi Ikeno}
\affil[1]{Department of Agricultural, Life and Environmental Sciences, Tottori University, Tottori 680-8551, Japan}

\author[2]{Takahiro Nishi}
\affil[2]{Nishina Center for Accelerator-Based Science, RIKEN, 2-1 Hirosawa,
Wako, Saitama 351-0198, Japan }

\author[2]{Kenta Itahashi}

\author[3]{Naoko Nose-Togawa}
\affil[3]{Research Center for Nuclear Physics (RCNP), Osaka University, Ibaraki 567-0047, Japan}%

\author[1]{Akari Tani}

\author[4]{Satoru Hirenzaki}
\affil[4]{Department of Physics, Nara Women's University, Nara 630-8506, Japan
\email{ikeno@tottori-u.ac.jp}}

\begin{abstract}
We investigate the possibility to determine the value of the pion--nucleon sigma term $\sigma_{\pi N}$ precisely by the experimental observables of the deeply bound pionic atoms. We discuss the sensitivity of the observables to $\sigma_{\pi N}$ and take into account the typical errors of the up-to-date experiments of the deeply bound pionic atoms. We find that the gap of the binding energies and the width of the deeply bound pionic states are good observables for the $\sigma_{\pi N}$ value determination by the experimental data.
We also discuss the expected difficulties for the accurate determination of the value of $\sigma_{\pi N}$ due to the correlation between the $\sigma$ term and the potential parameter in the pion--nucleus optical potential.
\end{abstract}

\maketitle


\section{Introduction}
Meson--Nucleus systems are known to provide valuable information on the meson properties at finite density~\cite{Batty:1997zp,Friedman:2007zza,Hayano:2008vn,Metag:2017yuh}. 
Especially, we think spectroscopic study of the deeply bound pionic atoms is very useful to investigate the pion properties~\cite{Piatom} and the aspects of the chiral symmetry at finite density~\cite{KSuzuki} based on the theoretical supports~\cite{Kolo,Jido}. 
The pion--nucleon sigma term $\sigma_{\pi N}$, which is defined as the nucleon matrix element of the mass terms of the light $u$ and $d$ quarks in QCD, is one of the essential quantities to investigate the value of the chiral condensate in the nuclear medium. The $\sigma_{\pi N}$ term is also important to know the contribution of the explicit chiral symmetry breaking to the nucleon mass. The value of $\sigma_{\pi N}$, however, has not been determined accurately enough. For example, from the compilation of the $\sigma_{\pi N}$ values~\cite{Yamanaka:2018uud,Gupta:2021ahb} we find that the studies based on the pion--nucleon scattering concluded $\sigma_{\pi N} \sim 60$~MeV~\cite{Alarcon:2011zs,Yao:2016vbz,Hoferichter:2015tha,RuizdeElvira:2017stg}, while the results by the lattice calculations seems to be distributed within the range of $\sigma_{\pi N}= 30 \sim 60$~MeV~\cite{Yamanaka:2018uud,Gupta:2021ahb,QCDSF-UKQCD:2011qop,Lutz:2014oxa,Durr:2015dna,Ren:2017fbv,Yang:2015uis,Abdel-Rehim:2016won,Bali:2016lvx,Alexandrou:2019brg}. The $\sigma_{\pi N}$ value determined in the previous analyses using the existing pionic atom data is reported to be $\sigma_{\pi N} = 57 \pm 7$~MeV~\cite{Friedman:2019zhc,Friedman:2020gsf} and is consistent with the results from the $\pi N$ scattering mentioned above. The analysis of the deeply bound pionic atoms in Ref.~\cite{KSuzuki}, on the other hand, indicates the $\sigma_{\pi N}$ value to be $\sigma_{\pi N} \sim 45$~MeV, which is in good agreement with Refs.~\cite{Gasser:1990ce,Gasser:1990ap}. 
In this exploratory level, we are much interested in the determination of the $\sigma_{\pi N}$ value by the precise data of the deeply bound pionic atoms expected to be obtained in near future~\cite{proposal2019}.
Actually the accuracy of each experimental datum is expected to be improved in coming experiments and we can also make use of the systematic data observed for the long Sn isotopes chain.

In the experimental studies of the deeply bound pionic atoms in Sn isotopes at the RI Beam Factory (RIBF), RIKEN~\cite{Nishi_Exp}, the formation spectra of the ($d,^3$He) reaction are measured so successfully that the angular dependence of the spectra is observed first time, and the binding energies and widths of the pionic 1$s$ and 2$p$ states are determined simultaneously with great accuracy.
So far, the deeply bound pionic atoms in tin were observed in $^{115, \ 119, \ 121, \ 123}$Sn isotopes~\cite{KSuzuki,Nishi_Exp}.  
Further experimental information with better precision is expected to be obtained for the pionic atoms in $^{111,123}$Sn by the ($d, ^3$He) reaction for $^{112,124}$Sn targets~\cite{proposal2019} and it might be very helpful for better determination of the $\sigma_{\pi N}$ term.

In this article, we investigate whether it is possible to determine the value of $\sigma_{\pi N}$ accurately by the deeply bound pionic atoms using the standard theoretical tools for the calculations of their structures and formation spectra. We discuss the sensitivity of the observables to $\sigma_{\pi N}$ and take into account the typical errors of the data  as partly reported in Ref.~\cite{Ikeno_Poland}.  We also discuss the expected difficulties to the accurate determination of the $\sigma_{\pi N}$ value. Then, we make some comments on the possible way for the precise determination of $\sigma_{\pi N}$.

\section{Formalism}\label{formalism}
We explain the standard theoretical formula here to investigate the structure and formation of the deeply bound pionic atoms, and study the $\sigma_{\pi N}$ term dependence of the deeply bound pionic atom observables by embedding $\sigma_{\pi N}$ into the potential parameters.

We solve the Klein--Gordon equation~\cite{Piatom,Ikeno},
\begin{equation}
\left[-\nabla^{2}+\mu^{2}+2\mu V_{\rm{opt}}(r)\right]\phi(\vec{r}\,)
=\left[ E- V_{\rm{em}}(r)\right]^{2} \phi(\vec{r}\,),
\label{KGeq}
\end{equation}
to study the structure of the pionic atoms, 
where $\mu$ is the pion--nucleus reduced mass, $E$ the eigen energy written as $E=\mu-B_{\pi}-\displaystyle \frac{i}{2}\Gamma_\pi$ with the binding energy $B_\pi$ and the width $\Gamma_\pi$ of the atomic states. 
The electromagnetic interaction $V_{\rm em}$ is described as~\cite{Ikeno:2015ioa,oset},
\begin{equation}
V_{{\rm em}}(r) = - \frac{e^2}{4 \pi \epsilon_0} \int \frac{\rho_{{\rm ch}}(r') Q ( | \vec{r} - \vec{r' }|) }{| \vec{r} - \vec{r' }|} d \vec{r '} , 
\label{eq:Vem}
\end{equation}
here $Q(r)$ is defined as, 
\begin{equation}
Q(r) = 1 + \frac{2}{3 \pi }\frac{e^2}{4 \pi \epsilon_0}\int^\infty_1 du\, e^{-2m_e ru}\left( 1+ \frac{1}{2u^2} \right)  \frac{(u^2-1)^{1/2}}{u^2} 
\end{equation}
with the electron mass $m_e$. This $V_{\rm em}$ includes the effects of the finite nuclear charge distribution $\rho_{\rm ch}(r)$ and the vacuum polarization.

\begin{table}[tb]
\caption{\label{table:Vopt_para}
Pion--nucleus optical potential parameters used in this article. The parameters are obtained in Ref~\cite{SM} for the so-called Ericson--Ericson potential~\cite{Ericson}. The parameters $b_0$ and $b_1$ are determined from the $\sigma_{\pi N}$ term as discussed in the text.
}
\centering
\begin{tabular}{ccc} 
\hline
 & Potential parameter  &~   \\ \hline
 & $b_{0}~ [m_{\pi}^{-1}]$ &~~~see text   \\
 & $b_{1}~ [m_{\pi}^{-1}]$ &~~~see text  \\	
 & $c_{0}~ [m_{\pi}^{-3}]$ &~~~ $  0.223$  \\
 & $c_{1}~ [m_{\pi}^{-3}]$ &~~~ $  0.25 $   \\	
 & $B_{0}~ [m_{\pi}^{-4}]$ &~~~ $ 0.042~i$  \\
 & $C_{0}~ [m_{\pi}^{-6}]$ &~~~ $ 0.10~i $  \\  
 & $\lambda$ &~~~ $ 1.0$  \\ 
\hline
\end{tabular}
\caption{\label{table:R,a}
The radius parameter $R_{\rm ch}$ of the charge distribution of the Sn isotopes~\cite{Fricke} used in Eq.~(\ref{rho_ch}).  The diffuseness parameter $a_{\rm ch}$ in Eq.~(\ref{rho_ch}) is fixed to be $a_{\rm ch}=t$/(4 ln3) for all nuclei with $t=2.30$ fm in Ref.~\cite{Fricke}. }
\centering
\begin{tabular}{cl|cccc} 
\hline
 &nuclide &~~ $^{112}$Sn &~~~~$^{124}$Sn~ & \\ \hline
 &$R_{\rm ch}$ [fm] &~~ 5.3714 &~~~~  5.4907 &~~~~ \\ 
\hline
\end{tabular}
\end{table}

We consider one of the standard optical potential, so-called 
Ericson--Ericson type~\cite{Ericson} written as,
\begin{eqnarray}
2\mu V_{\rm opt}(r)
=-4\pi[b(r)+\varepsilon_2B_0\rho^2(r)] 
 +4\pi\nabla\cdot[c(r)+\varepsilon_2^{-1}C_0\rho^2(r)]L(r)\nabla,
\label{eq:Vopt}
\end{eqnarray}
with
\begin{eqnarray}
b(r) &=& \varepsilon_{1}[b_{0}\rho(r)+b_{1}[\rho_n(r)-\rho_p(r)]], \label{Vopt_Swave1}\\
c(r) &=& \varepsilon_{1}^{-1}[c_{0}\rho(r)+c_{1} [\rho_n(r)-\rho_p(r)]],\label{Vopt_Pwave2}\\
L(r) &=& \left\{ 1+ \frac{4}{3} \pi\lambda[c(r)+
      \varepsilon_{2}^{-1}C_{0}\rho^{2} (r)] \right\}^{-1}, \label{Vopt_Pwave3}
\end{eqnarray}
where $\varepsilon_{1}$ and $\varepsilon_{2}$ are defined as
$\varepsilon_{1}=1+\displaystyle \frac{\mu}{M}$ and 
$\varepsilon_{2}=1+\displaystyle \frac{\mu}{2M}$ with 
the nucleon mass $M$.  
The parameters $b$’s and $c$’s indicate the $s$-wave and $p$-wave $\pi N$ interaction, respectively. The parameters $b_0$ and $b_1$ are replaced by the density dependent form with the $\sigma_{\pi N}$ term as explained later.
The potential terms with parameter $B_0$ and $C_0$ are higher order contributions to the optical potential, and $\lambda$ the Lorentz-Lorenz correction.
We use in this article the potential parameters obtained in Ref.~\cite{SM} except for $b_0$ and $b_1$, which are compiled in Table~\ref{table:Vopt_para}.

As for the nuclear densities appeared in the electromagnetic interaction $V_{\rm em}$ and the pion--nucleus optical potential $V_{\rm opt}$, we use the Woods--Saxon form.
The charge density distribution $\rho_{\rm ch}$ in Eq.~(\ref{eq:Vem}), which is normalized to the nuclear charge, is written as,
\begin{equation}
\rho_{\rm ch}(r) = \frac{ \rho_{\rm ch 0}}{1+ \exp[(r-R_{\rm ch})/a_{\rm ch}]},
\label{rho_ch}
\end{equation}
with the radius parameter $R_{\rm ch}$ and the diffuseness parameter $a_{\rm ch}$.
The values of the radius parameter $R_{\rm ch}$ and the diffuseness parameter $a_{\rm ch}$ are taken from Ref.~\cite{Fricke} and shown in Table~\ref{table:R,a}.
The parameter $a_{\rm ch}$ in Ref.~\cite{Fricke} is fixed to be $a_{\rm ch}=t$/(4 ln3) for all nuclei with $t=2.30$ fm.

The distributions of the center of nucleon ($\rho$), proton ($\rho_p$), and neutron ($\rho_n$) appeared in Eqs.~(\ref{eq:Vopt})--(\ref{Vopt_Pwave3}) are also written by the Woods-Saxon form as,
\begin{equation}
\rho(r)=\rho_{p}(r)+ \rho _{n}(r)
=\frac{ \rho_{0}}{1+ \exp[(r-R)/a]},
\label{rho}
\end{equation}
where we assume the same distribution shape for the both of the proton and the neutron distributions.
The densities $\rho$, $\rho_p$ and $\rho_n$ are normalized to be mass, proton and neutron numbers, respectively.
The radius and diffuseness parameters $R$ and $a$ are determined from the parameters $R_{\rm ch}$ and $a_{\rm ch}$ of the charge distribution $\rho_{\rm ch}$ by the prescription described in Ref.~\cite{oset}.

We consider the pion--nucleus optical potentials in which the $\sigma_{\pi N}$ term is embedded to study the sensitivities of the observables of the deeply bound pionic atoms to the value of the $\sigma_{\pi N}$ term. We follow the form proposed in Refs.~\cite{Weise:2000xp,Weise:2001sg} based on the Tomozawa~\cite{Tomozawa:1966jm}--Weinberg~\cite{Weinberg:1966kf} and the Gell-Mann--Oakes--Renner~\cite{Gell-Mann:1968hlm} relations, and determine the value of the $s$-wave isovector potential parameter $b_1$ in terms of $\sigma_{\pi N}$ as,
\begin{equation} 
b_1(\rho) =b_1^{\rm free}\left(1-\frac{\sigma_{\pi N}}{m_{\pi}^2f_{\pi}^2} \rho\right)^{-1},
\label{eq:b1_sigma} 
\end{equation} 
where $b_1^{\rm free}$ is the isovector  $\pi N$ scattering length in vacuum $b_1^{\rm free}=-0.0861 \ m_{\pi}^{-1}$~\cite{Friedman:2019zhc,Friedman:2020gsf,Baru:2010xn}, and $f_\pi$ the pion decay constant in vacuum $f_{\pi}=92.4$~MeV~\cite{Weise:2000xp}.
This form includes the lowest order term only.
We adopt this form in this article 
as the simple and robust form to connect the potential parameters to the $\sigma_{\pi N}$ term, and to study and clarify the sensitivities of the observables of the deeply bound pionic atoms to the $\sigma_{\pi N}$ value.
The same form is also adopted in Refs.~\cite{Friedman:2019zhc,Friedman:2020gsf} for the $\sigma_{\pi N}$ value determination by the data of the existing pionic atoms mainly bounded in the light nuclei.   
For the actual determination of the $\sigma_{\pi N}$ value, more sophisticated theoretical formula would be necessary.

As indicated in Eq.~(\ref{eq:b1_sigma}), the $b_1$ parameter in the potential has the explicit density dependence by including the $\sigma_{\pi N}$ term. We, then, also take into account the double scattering effects to the $s$-wave isoscalar potential parameter $b_0$~\cite{Ericson} with the density dependent $b_1$ parameter in Eq.~(\ref{eq:b1_sigma}) as,
\begin{eqnarray}
b_0(\rho) =  b_0^{\rm free} - \varepsilon_1 \frac{3}{2\pi} ( b_0^{\rm free \, 2} + 2b_1^2(\rho) ) \left( \frac{3\pi^2}{2} \rho \right)^{1/3},
\label{eq:b0_double} 
\end{eqnarray}
where $b_0^{\rm free}$ is the isoscalar $\pi N$ scattering length in vacuum $b_0^{\rm free}=0.0076 \ m_{\pi}^{-1}$~\cite{Friedman:2019zhc,Friedman:2020gsf,Baru:2010xn}. In Eq.~(\ref{eq:b0_double}), the local Fermi momentum of the nucleon is expressed by the nuclear density $\rho$ as $ \displaystyle \left( \frac{3\pi^2}{2} \rho \right)^{1/3}$.
Thus, the explicit $\sigma_{\pi N}$ term inclusion requires to consider the density dependent $b_0$ and $b_1$ parameters in the optical potential.

We also calculate the pionic atom formation spectra in the ($d, ^3$He) reaction with the effective number approach~\cite{Ikeno:2015ioa,Ikeno:2011aa,Umemoto}.
The spectra at the forward angles for the Sn isotopes have been calculated in Ref.~\cite{Umemoto,Umemoto_Dthesis} and at the finite angles in Refs.~\cite{Ikeno:2015ioa,Ikeno:2011aa}. We follow the same formula in this article to calculate the formation spectra.

\section{Results and Discussions}\label{result}
In this section, we show the calculated results of the observables of the deeply bound  pionic atoms with the different $\sigma_{\pi N}$ values within the range of $25 \leq \sigma_{\pi N} \leq 60$~MeV to study the sensitivities of them to $\sigma_{\pi N}$.

\begin{figure}[tb]
\centering
 \includegraphics[scale=0.73]{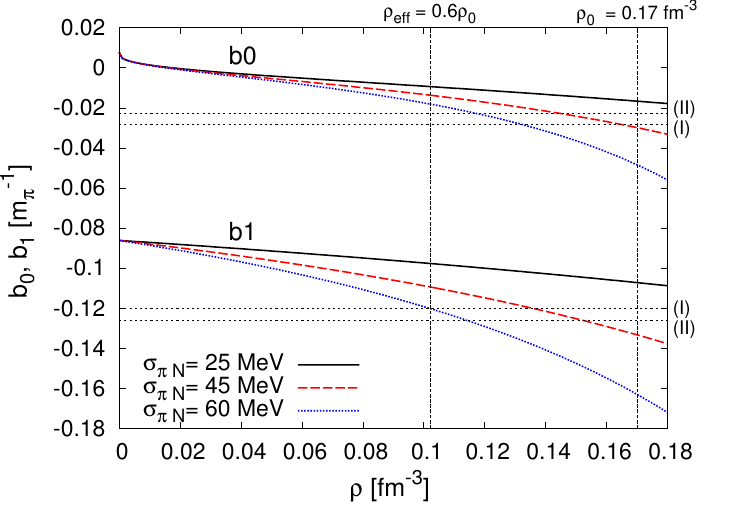}
 \caption{Density dependence of the $b_0(\rho)$ (Eq.~(\ref{eq:b0_double})) and $b_1(\rho)$ (Eq.~(\ref{eq:b1_sigma})) parameters are shown for the different $\sigma_{\pi N}$ values as indicated in the figure.
The values of the constant $b_0$ and $b_1$ parameters in Refs.~\cite{SM} and \cite{Exp_itahashi} indicated  as (I) and (II) are also shown by the dotted lines. The normal nuclear density $\rho_0$ and the effective density $\rho_{\rm eff}$~\cite{rhoe} are also indicated in the figure.
  }
 \label{fig:b0b1}
\end{figure}

In Fig.~\ref{fig:b0b1}, first we show the density dependence of the $b_0(\rho)$ parameter defined in Eq.~(\ref{eq:b0_double}) and the $b_1(\rho)$ parameter in Eq.~(\ref{eq:b1_sigma}) for three different $\sigma_{\pi N}$ values, $\sigma_{\pi N} = 25$, 45, and 60~MeV.
The normal nuclear density $\rho_0 = 0.17$~fm$^{-3}$ and the effective density $\rho_{\rm eff} = 0.6~\rho_0$ for the pionic atoms, which is introduced in Ref.~\cite{rhoe} and used in the analysis of Ref.~\cite{KSuzuki}, are indicated in the figure. In Fig.~\ref{fig:b0b1}, we also plot the constant $b_0$ and $b_1$ parameter values obtained in Refs.~\cite{SM} and \cite{Exp_itahashi} for comparison.
The larger $\sigma_{\pi N}$ value causes the stronger density dependence of the parameters and provides the more repulsive pion--nucleus $s$-wave interaction. We find that the $b_1(\rho)$ value at the effective density $\rho_{\rm eff}$ is almost consistent with the constant $b_1$ value in Refs.~\cite{SM} for the value of $\sigma_{\pi N} = 60$~MeV.

\begin{figure}[tb]
\centering
\includegraphics[scale=0.68]{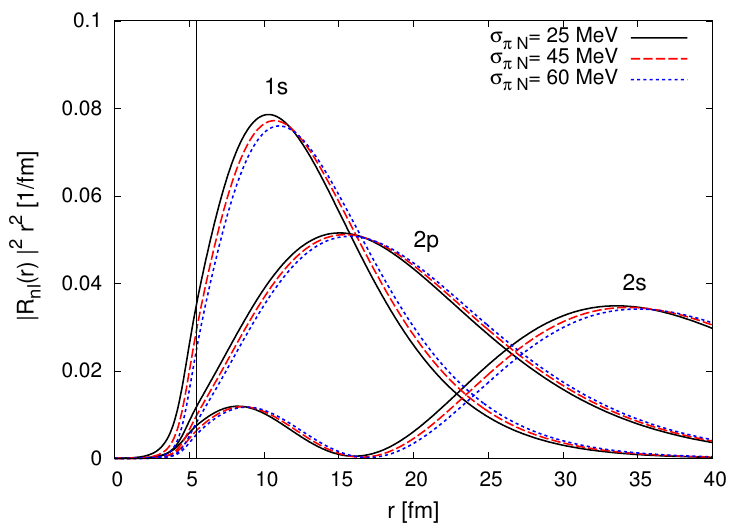}
 \caption{The radial density distributions of the pionic 1$s$, 2$p$, and 2$s$ states in $^{123}$Sn are plotted as the functions of the radial coordinate $r$
for the different $\sigma_{\pi N}$ values as indicated in the figure. The density dependent $b_0(\rho)$ and $b_1(\rho)$ parameters are used. The vertical line shows the radius of $^{123}$Sn.}
 \label{fig:wf}
\end{figure}

The structures of the deeply bound states are obtained by solving the Klein--Gordon equation with the optical potential Eqs.~(\ref{eq:Vopt})--(\ref{Vopt_Pwave3}). We use the density dependent $b_0(\rho)$ and $b_1(\rho)$ instead of the constant $b_0$ and $b_1$ values and study how the structure of the states changes with the different values of the $\sigma_{\pi N}$ term. We show in Fig.~\ref{fig:wf} the calculated pionic radial density distributions in $^{123}$Sn with the $b_0(\rho)$ and $b_1(\rho)$ parameters for $\sigma_{\pi N} = 25$, 45, and 60~MeV cases. We can see from the figures that the densities are pushed more outwards for the larger $\sigma_{\pi N}$ values because of the stronger repulsive effects of the potential.

\begin{figure}[tb]
\centering
\includegraphics[scale=0.7]{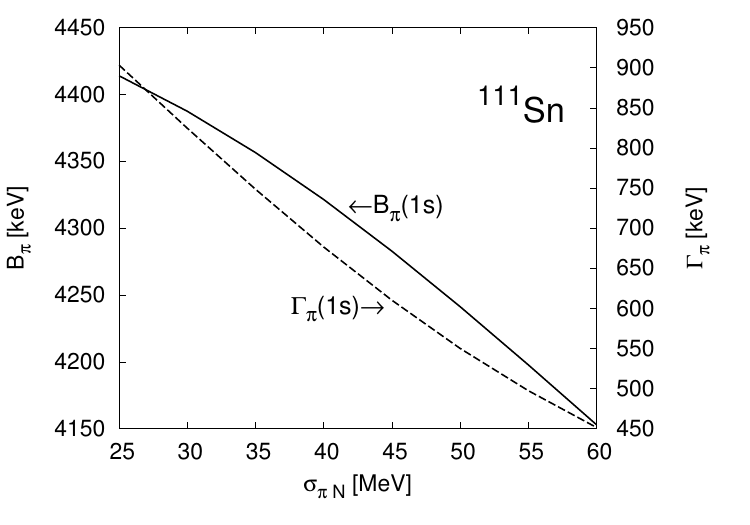} \\
 \caption{The binding energy ($B_\pi$) and the width ($\Gamma_\pi$) of the pionic 1$s$ state in $^{111}$Sn are plotted as the functions of the $\sigma_{\pi N}$ value. $B_\pi$ and $\Gamma_\pi$ are calculated with the density dependent $b_0(\rho)$ and $b_1(\rho)$ parameters.
}
 \label{fig:BE_sigma}
\end{figure}

\begin{table*}[htb]
\caption{\label{table:BE}
The calculated average shifts of the observables of the deeply bound pionic states are shown in the unit of keV for the 1~MeV change of the $\sigma_{\pi N}$ value $\Delta \sigma_{\pi N}=1$~MeV. $\Delta (B_\pi(1s)- B_\pi(2p) )$ and $\Delta ( \Gamma_\pi(1s) - \Gamma_\pi(2p) ) $ indicate the average shifts of the differences of the binding energies and widths between the $1s$ and $2p$ states for the $\sigma_{\pi N}$ change $\Delta \sigma_{\pi N} = 1$~MeV, respectively.
 }
\centering
\begin{tabular}{cc|c|c} 
\hline
 & $\left[ {\rm keV} \right]$~~~~ &~~~~~ $^{123}$Sn ~~~~~ &~~~~~$^{111}$Sn~~~~~ \\ \hline
 & $ \left| \Delta B_\pi(1s) \right| $ &~~ 6.2  &~~~~ 7.5 \\ 
 & $ \left| \Delta \Gamma_\pi(1s) \right| $  &~~ 5.9  &~~~~ 12.9 \\ \hline
 & $ \left| \Delta B_\pi(2p)  \right| $ &~~ 1.7  &~~~~ 1.7 \\ 
 & $ \left| \Delta \Gamma_\pi(2p)  \right| $  &~~ 2.5   &~~~~3.6 \\ \hline
 & $ |\Delta (B_\pi(1s)- B_\pi(2p) )| $ &~~ 4.5  &~~~~ 5.8 \\ 
 & $ |\Delta ( \Gamma_\pi(1s) - \Gamma_\pi(2p) )| $  &~~ 3.4    &~~~ 9.3 \\ 
\hline
\end{tabular}
\end{table*}

\begin{figure}[htb]
\centering
 \includegraphics[scale=0.7]{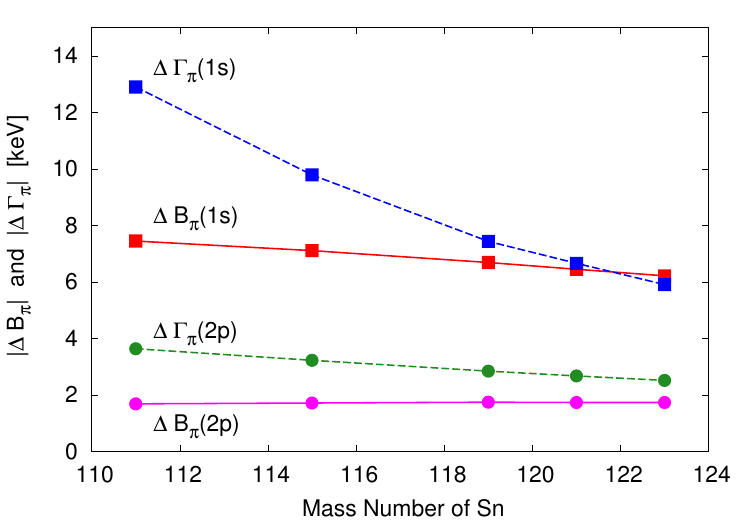} 
 \caption{ The mass number dependence of the calculated shifts of the observables of the deeply bound pionic states is shown for the 1~MeV change of the $\sigma_{\pi N}$ value $\Delta \sigma_{\pi N}=1$~MeV in Sn isotopes.
}
 \label{fig:BE_sigma_Sn}
\end{figure}

In Fig.~\ref{fig:BE_sigma}, the binding energy and the width of the deeply bound 1$s$ state in $^{111}$Sn are plotted as the functions of the $\sigma_{\pi N}$ value. 
In the figure, we find that each observable depends on the $\sigma_{\pi N}$ value almost linearly within the range of the $\sigma_{\pi N}$ value considered here. Thus, we use the average slope of the line, namely the average size of the shift of each observable due to the 1~MeV variation of the $\sigma_{\pi N}$ value $\Delta \sigma_{\pi N} =1$~MeV, to express the sensitivity of the observable to $\sigma_{\pi N}$.
We note here that this figure should not be used directly to determine the $\sigma_{\pi N}$ value by the binding energy and/or the width since the figure is just to show the sensitivities of the observables to the $\sigma_{\pi N}$ value. To determine the absolute value of $\sigma_{\pi N}$, we need thorough analyses of the data in general.
The sizes of the calculated sensitivity of the observables are compiled in Table~\ref{table:BE} for the cases considered in this article.
We find that the sensitivities of the $1s$ state observables are stronger than those of the $2p$ states as naturally expected and the shifts of the $1s$ state binding energy $\Delta B_\pi (1s)$ is $|\Delta B_\pi (1s)| = 6.2$~keV for $^{123}$Sn and 7.5~keV for $^{111}$Sn for the 1~MeV variation of the $\sigma_{\pi N}$ value $\Delta \sigma_{\pi N} = 1$~MeV.  
We find the larger sensitivities for the pionic states in lighter Sn isotope $^{111}$Sn to $\sigma_{\pi N}$ because of the less repulsive optical potential due to the smaller neutron numbers and the larger overlap of pionic wave function with nucleus.
The shift of the width of the $1s$ pionic states $\Delta \Gamma_\pi(1s)$ in $^{111}$Sn is 12.9~keV for the $\Delta \sigma_\pi = 1$~MeV variation, which is more than twice of $\Delta \Gamma_\pi(1s)$ in $^{123}$Sn cases as shown in Table~\ref{table:BE}. We show the mass number dependence of the sensitivity of each observable in Fig.~\ref{fig:BE_sigma_Sn} and find clearly the stronger sensitivities of the observables, especially $\Gamma_\pi (1s)$, for lighter Sn isotopes.

These calculated sensitivities of the observables can be compared with the accuracy of the latest experimental data~\cite{Nishi_Exp,Exp_itahashi}.
The typical errors of the up-to-date experiments for the deeply bound pionic atom observables by the ($d,^3$He) reactions in Sn region are around 80~keV for the binding energy of the $1s$ state and around 40~keV for the width of the $1s$ state. Some of the combinations of the observables are known to have the advantages to reduce the systematic errors. For our purpose, the gap of the binding energies of the $1s$ and $2p$ states, $B_\pi(1s) - B_\pi(2p)$, is considered to be important since they can be determined far more accurately and its error is expected to be 10 $\sim$ 15~keV for Sn region.
We can estimate the uncertainties of the $\sigma_{\pi N}$ value determination by the expected experimental errors and the calculated sensitivities of the observables.
The calculated sensitivity of the energy gap $|\Delta (B_\pi(1s) - B_\pi(2p))|$ for $^{111}$Sn is 5.8~keV as shown in Table~\ref{table:BE}. In this case, the experimental error 10 $\sim$ 15~keV of this energy gap can be interpreted as the uncertainty of the $\sigma_{\pi N}$ value $\displaystyle 1.7\sim 2.6$~MeV, which is obtained by dividing the experimental error 10 $\sim$ 15~keV by the sensitivity of the observable 5.8~keV for the 1~MeV change of the $\sigma_{\pi N}$ value.  
The sensitivities of the energy gap obtained here are 4.5~keV for $^{123}$Sn and 5.8~keV for $^{111}$Sn as shown in Table~\ref{table:BE}. The corresponding uncertainties of $\sigma_{\pi N}$ to the experimental error 10 $\sim$ 15~keV, thus, distribute within the range of 1.7 $\sim$ 3.3~MeV. Hence, we evaluate the uncertainty of the $\sigma_{\pi N}$ value determination using the data of $ B_\pi (1s) - B_\pi(2p) $ to be around 2.5~MeV.
This is much better than the uncertainties based on the use of the absolute value of $B_\pi(1s)$ for $^{111}$Sn. The expected size of the experimental error 80~keV of $B_\pi(1s)$ and the calculated sensitivity 7.5~keV for the 1~MeV change of $\sigma_{\pi N}$ in $^{111}$Sn conclude that the expected uncertainty of the $\sigma_{\pi N}$ value is large and would be $\displaystyle 11$~MeV which is estimated as 80~keV/(7.5~keV/$\Delta \sigma =1$~MeV). While, the width of the $1s$ state for $^{111}$Sn provides the relatively small expected uncertainty of the $\sigma_{\pi N}$ value to be $\displaystyle 3.1$~MeV for the experimental error 40~keV.
Thus, we find from the typical size of the experimental errors and the calculated sensitivities of the observables that the energy gap between the $1s$ and $2p$ states, and the width of the $1s$ state in lighter Sn isotopes have the larger possibility to provide the important information to determine the $\sigma_{\pi N}$ value precisely.

\begin{figure}[tb]
\centering
\includegraphics[scale=0.55]{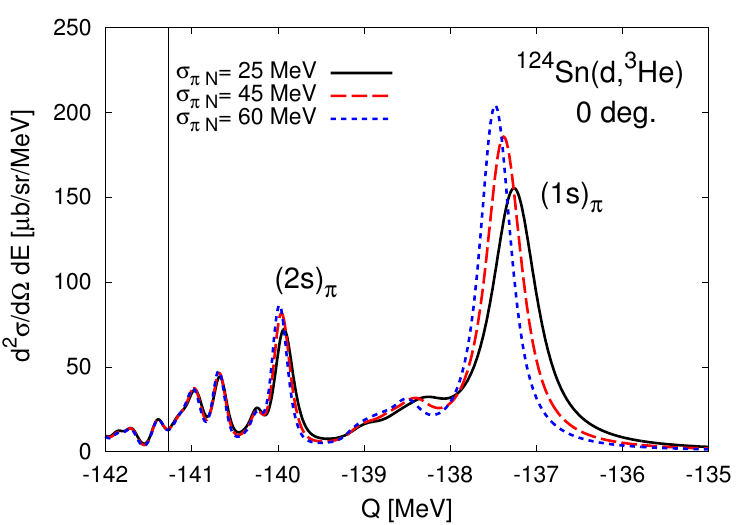}
\includegraphics[scale=0.55]{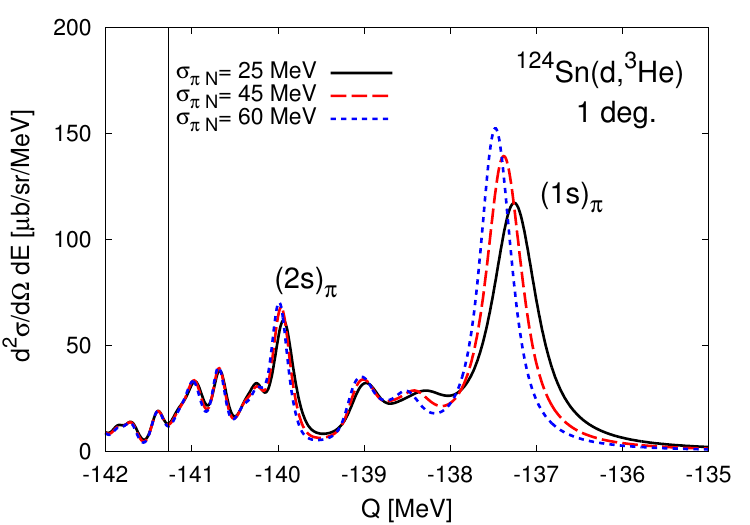}
\includegraphics[scale=0.55]{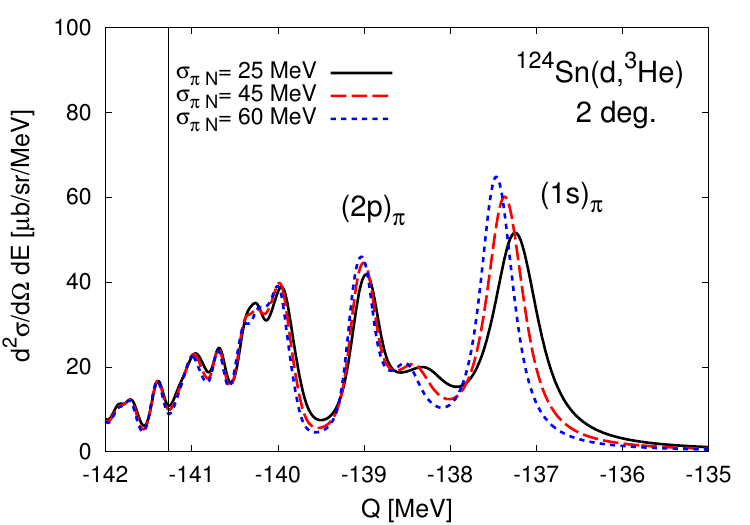}
\caption{Formation cross sections of the deeply bound pionic atoms in $^{123}$Sn by the  $^{124}$Sn($d, ^3$He) reactions are shown at the different scattering angles of the emitted $^{3}$He nucleus in the laboratory frame as $\theta_{\rm d He}^{\rm Lab} = 0^\circ$, $1^\circ$,  $2^\circ$, respectively. The results are obtained with the density dependent $b_0(\rho)$ and $b_1(\rho)$ parameters with three different $\sigma_{\pi N}$ values as indicated in the figure. Experimental energy resolution is assumed to be $\Delta E = 150$~keV.
The contributions from the quasi-free pion production are not included in the theoretical spectra.
}
\label{fig:cross_124Sn}
\end{figure}

\begin{figure}[tb]
\centering
\includegraphics[scale=0.55]{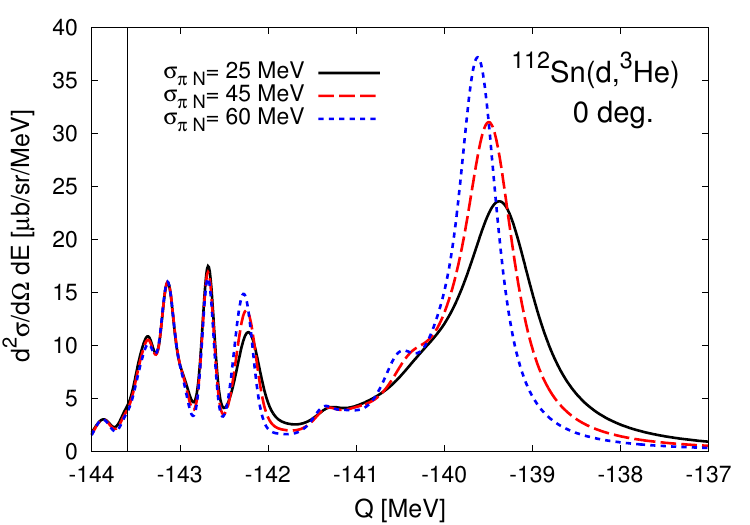}
\includegraphics[scale=0.55]{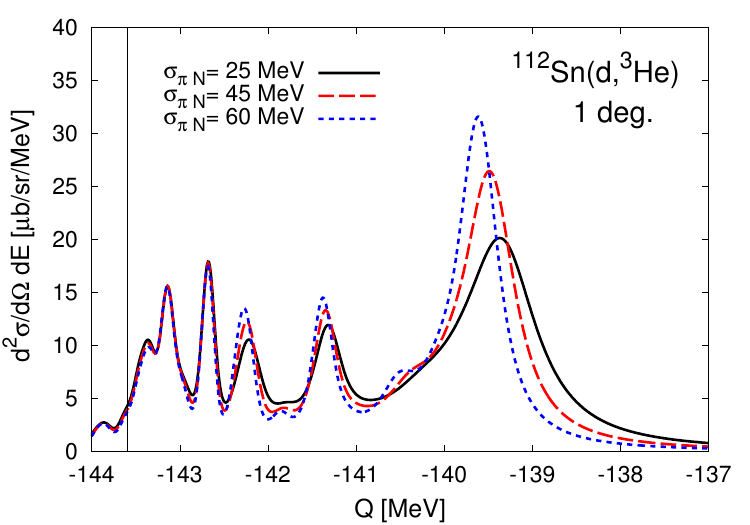}
\includegraphics[scale=0.55]{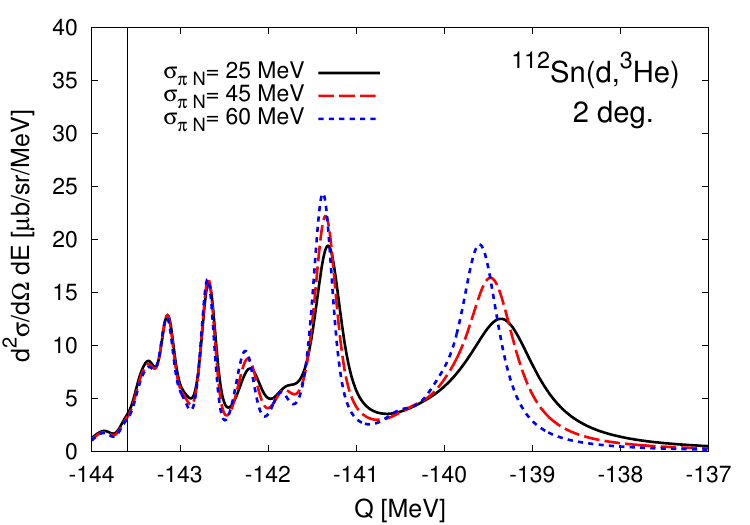}
\caption{Same as Fig.~\ref{fig:cross_124Sn} except for the deeply bound pionic atoms in $^{111}$Sn by the $^{112}$Sn($d, ^3$He) reactions.}
\label{fig:cross_112Sn}
\end{figure}

Then, we show in Figs.~\ref{fig:cross_124Sn} and \ref{fig:cross_112Sn} the calculated spectra of the ($d, ^3$He) reactions for the formation of the deeply bound pionic states for $^{112, 124}$Sn targets obtained with the density dependent $b_0(\rho)$ and $b_1(\rho)$ parameters. We find that the shape of the spectrum have the reasonable sensitivity to the $\sigma_{\pi N}$ value at each scattering angle. Especially, the peak height of the pionic 1$s$ state formation is clearly reduced for the smaller $\sigma_{\pi N}$ values for all cases considered here because of the less repulsive potential and thus, larger absorptive width of the state.
In the detailed analyses of the shape of the experimental formation spectra with the theoretical results, we can expect to obtain the information on the potential parameters including the $\sigma_{\pi N}$ value~\cite{Nishi_Exp,Exp_itahashi}. 
As a possibility, the behavior of the shape of the tail of the largest peak structure due to 1$s$ bound state formation could provide the extra information on the $\sigma_{\pi N}$ value in addition to those from the binding energy and width of the state.

\begin{figure}[tb!]
\centering
\includegraphics[scale=0.65]{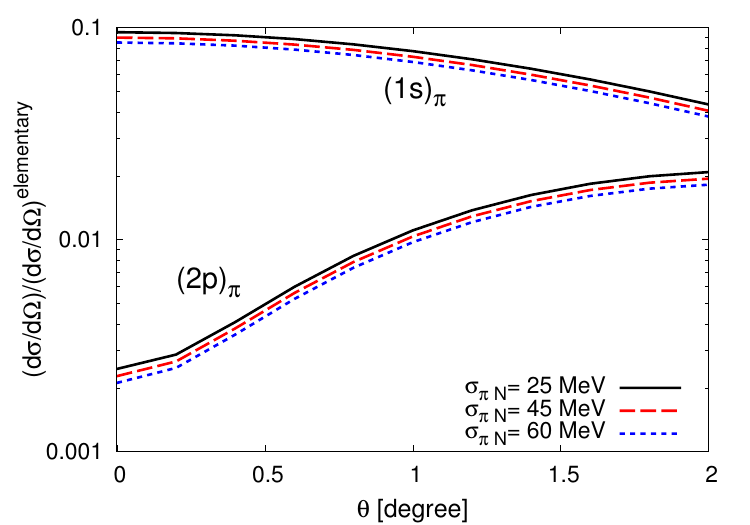}
\includegraphics[scale=0.65]{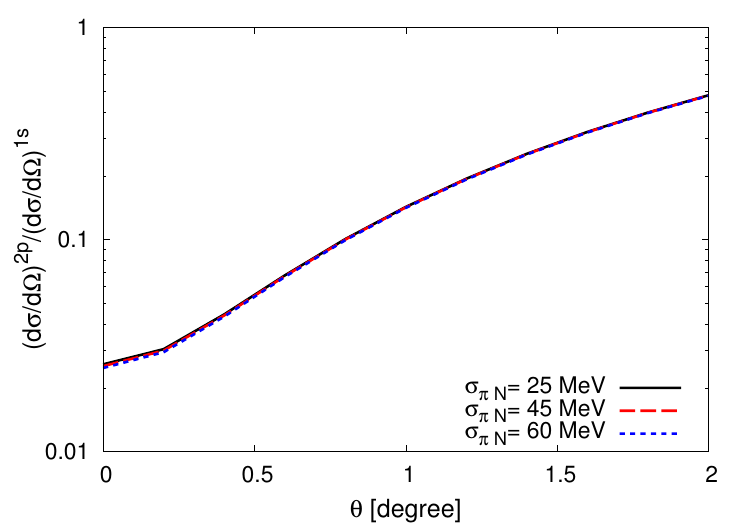}
\caption{ 
(Upper) The calculated angular dependence of the differential cross sections $\displaystyle \left( \frac{d \sigma}{d \Omega} \right)$ for the pionic 1$s$ and 2$p$ state formation in the $^{124}$Sn($d, ^3$He) reaction is shown in the unit of the elementary cross section $\displaystyle \left( \frac{d \sigma}{d \Omega} \right)^{\rm elementary}$. The cross sections are obtained as the sum of the contributions of all neutron hole states, and are calculated  with the density dependent $b_0(\rho)$ and $b_1(\rho)$ parameters with three different $\sigma_{\pi N}$ values as indicated in the figure.
(Lower) The angular dependence of the ratio of the pionic 1$s$ and 2$p$ states formation in the $^{124}$Sn($d, ^3$He) reaction with the density dependent $b_0(\rho)$ and $b_1(\rho)$.
}
 \label{fig:cross_124Sn_SM_angle}
\end{figure}

We show in Fig.~\ref{fig:cross_124Sn_SM_angle} (Upper) the calculated angular dependence of the differential cross sections $\displaystyle \left( \frac{d \sigma}{d \Omega} \right)$ for the pionic 1$s$ and 2$p$ state formation in the unit of the cross section of the elementary process $d + n \to {^3 \rm He} + \pi^-$. The differential cross sections for the specific pionic state formation are obtained theoretically by summing up the contributions of all neutron hole states. We also show the angular dependence of the ratio of the 1$s$ and 2$p$ state formation in Fig.~\ref{fig:cross_124Sn_SM_angle} (Lower) as the similar plot of the experimental data in Fig.~4 in Ref.~\cite{Nishi_Exp}. We find that the angular dependence of the formation cross section is quite stable to the change of the $\sigma_{\pi N}$ value and has only rather weak sensitivities to the $\sigma_{\pi N}$ value.

Here, we discuss the expected difficulties for the actual determination of the value of the $\sigma_{\pi N}$ term from the experimental data. First of all, it is well known that there exists the 
strong correlation between potential parameters $b_0$ and Re$B_0$~\cite{SM}.  
In the present form shown in Eqs.~(\ref{eq:b1_sigma}) and (\ref{eq:b0_double}), parameters $b_0$ and $b_1$ are both connected to the $\sigma_{\pi N}$ value. Thus, there could be a strong correlation between $\sigma_{\pi N}$ and Re$B_0$ which implies that $b_0$ and $b_1$ are both strongly correlated to Re$B_0$.

\begin{figure*}[tb!]
\centering
\includegraphics[scale=0.66]{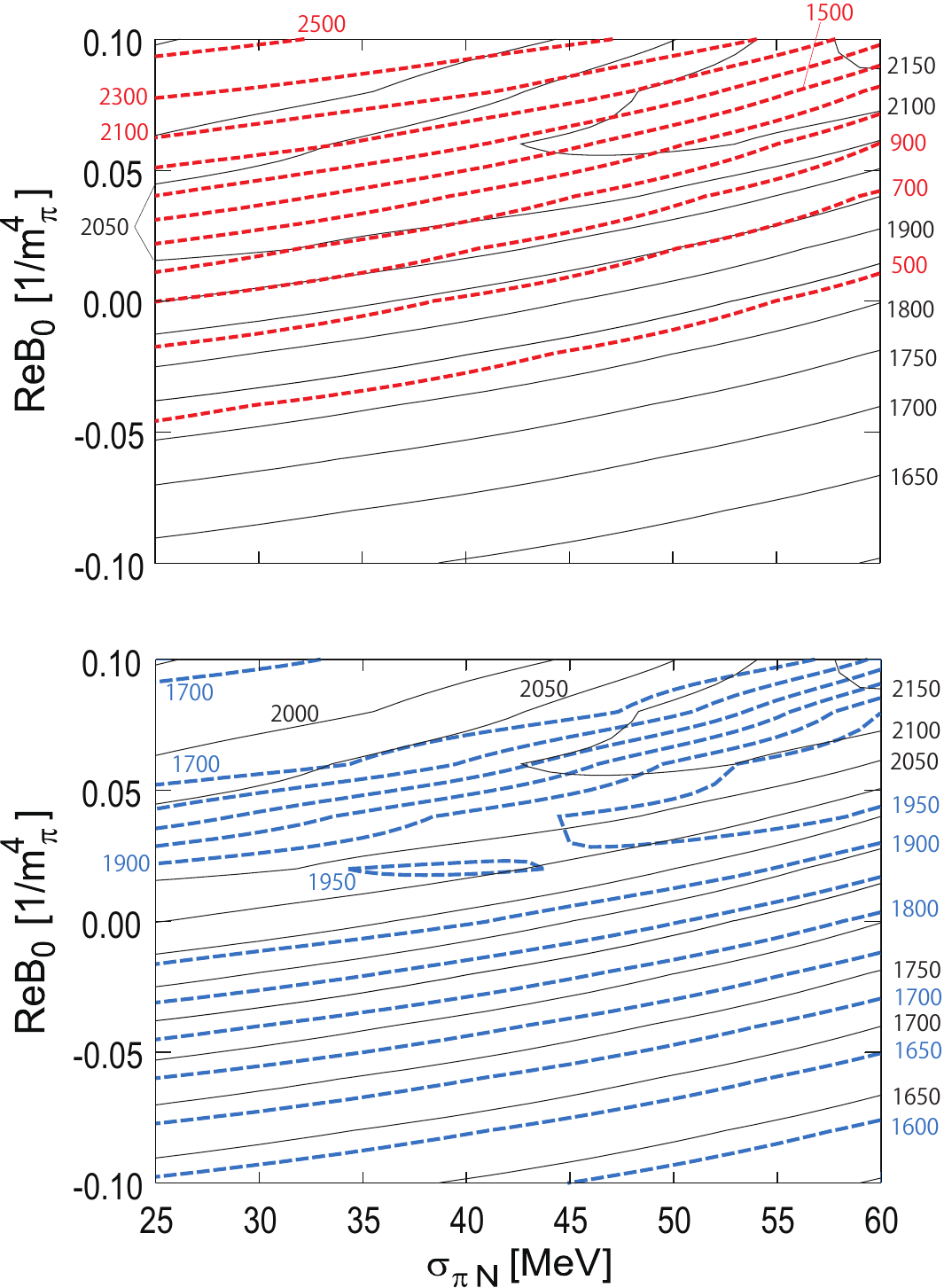}
\caption{ The contour plots of the observables 
 of the deeply bound pionic atoms in $^{111}$Sn in the $\sigma_{\pi N}-\mathrm{Re}B_0$ plane. The numbers written in the contour plots are in the unit of keV. (Upper) The solid and the dashed lines show the contour plots of 
the gap of the binding energies of the 1$s$ and 2$p$ states $B_\pi (1s)- B_\pi(2p)$ 
and 
the width of the 1$s$ state $\Gamma_\pi(1s)$
 calculated with the same shape of the proton ($\rho_p$) and the neutron ($\rho_n$) distributions with the appropriate normalization as described in the text. 
(Lower) The solid lines are the same as those in the upper figure. The dashed lines show $B_\pi (1s)- B_\pi(2p)$ for the nuclear distributions with $R_p = R$ and $R_n = R+0.2$~fm where $R_p$ and $R_n$ are the radius parameters of $\rho_p$ and $\rho_n$  written in the Woods-Saxon form. The parameter $R$ is same as in Eq.~(\ref{rho}) in the text. Between two dashed lines of 1700~keV in the positive Re$B_0$ region, there exists the shallow valley structure around 1675~keV.
} 
 \label{fig:contour}
\end{figure*}

We show in Fig.~\ref{fig:contour} the contour plots of the observables of the deeply bound pionic atoms in the plane of the sigma term $\sigma_{\pi N}$ and the potential parameter Re$B_0$. In Fig.~\ref{fig:contour} (Upper), we show the gap of the binding energies $B_\pi (1s)- B_\pi(2p)$ and the width $\Gamma_\pi(1s)$ for $^{111}$Sn to see the correlations of the parameters for those observables, and to study the possibilities to determine the individual parameter by the precise measurements of the binding energies and widths of the specific nucleus. We find that the correlations of $\sigma_{\pi N}$ and Re$B_0$ for the two observables looks more similar in the Re$B_0 <0$ region and it would be difficult to determine each parameter only by these observables. In the Re$B_0 > 0$ region, we find the correlations of these parameters show different patterns for these observables and we expect to have larger possibilities for the determination of the parameters.

In Fig.~\ref{fig:contour} (Lower), we show the contour plots of $B_\pi (1s)- B_\pi(2p)$ for the nuclei with the different neutron densities. We change the radius parameter of the neutron distribution as $R_n \to R_n + 0.2$~fm to simulate the effects of the neutron skin. We find again the relatively different parameter correlations in the Re$B_0 > 0$  region and larger possibilities to determine the parameters by the observables of the deeply bound pionic atoms in the nuclei with the different neutron densities, while in the Re$B_0 < 0$ region, the contour lines are almost parallel and it seems more difficult to determine both $\sigma_{\pi N}$ and Re$B_0$ precisely.

We also made the similar studies for the 1$s$ and 2$p$ pionic states in $^{205}$Pb which were observed by the $^{206}$Pb($d,^3$He) reaction~\cite{Hirenzaki:1997at,Geissel:2002ur}, and found the similar results as in the $^{111}$Sn cases described above.
Thus, we need to anticipate the difficulties by the parameter correlations to determine precisely both Re$B_0$ and $\sigma_{\pi N}$ simultaneously.  

Another expected difficulty is the neutron distribution of the nucleus, which is very important to obtain the potential parameters, especially for isovector $b_1$ parameter and thus for $\sigma_{\pi N}$. 
The neutron distributions are not well determined generally, and in Refs.~\cite{Friedman:2019zhc,Friedman:2020gsf}, the densities of the neutron are also included in the fitting procedure with the optical potential parameters by the pionic atom data.  We think it is better to use the accurate neutron densities determined by the independent experiments. For example, we can use the densities obtained in Ref.~\cite{Terashima:2008zza} for the analyses of the deeply bound pionic atoms in Sn, and could proceed to determine the potential parameters including the $\sigma_{\pi N}$ value.  
In order to obtain the $\sigma_{\pi N}$ value precisely in spite of the parameter correlations mentioned above, we think we need the combined analyses of the deeply bound pionic atoms in different nuclei for which the neutron densities are determined accurately by the independent experiments.

\section{Conclusions}\label{conclusion}
In this article, we study the sensitivities of the observables of the deeply bound pionic atoms to the value of the pion--nucleon sigma term $\sigma_{\pi N}$ and investigate the experimental feasibilities of them to determine the $\sigma_{\pi N}$ value precisely by taking into account the typical errors of the up-to-date experiments.
So far, the analyses of the data have been performed based on the usage of the effective nuclear density $\rho_{\rm eff}$ probed by the pionic atoms which would be slightly different for different nuclei and different bound states. 
In this article, we improve the theoretical formula and implement the $\sigma_{\pi N}$ term in the optical potential to treat the density dependence of the potential parameters for the $s$-wave isoscalar ($b_0$) and isovector ($b_1$) terms explicitly without using the concept of the effective density.

We calculate the various observables and study the sensitivities of them to the $\sigma_{\pi N}$ value for the deeply bound pionic atoms mainly in $^{111}$Sn and $^{123}$Sn.
We find that the binding energies and widths of the pionic $1s$ states have the largest sensitivities to the $\sigma_{\pi N}$ value. The sensitivities tend to be even larger for the lighter Sn isotopes, and the shifts of the $1s$ binding energy $\Delta B_\pi (1s)$ and the $1s$ width $\Delta \Gamma_\pi (1s)$ in $^{111}$Sn are found to be $\Delta B_\pi (1s)= 7.5$~keV and $\Delta \Gamma_\pi (1s)=12.9$~keV for the variation of the value of the $\sigma_{\pi N}$ term $\Delta \sigma_{\pi N} =1$~MeV. By considering the expected errors of the up-to-date experiments, we conclude that the energy gap of the $1s$ and $2p$ pionic states $B_\pi(1s) - B_\pi(2p) $ and the width of the $1s$ state for the lighter Sn isotope are expected to be most important observables to determine the $\sigma_{\pi N}$ value precisely. The uncertainties to the $\sigma_{\pi N}$ value due to the experimental errors to these observables are estimated to be 2.5~MeV for the energy gap $B_\pi(1s) - B_\pi(2p)$ for $^{123}$Sn and $^{111}$Sn, and 3.1~MeV for the width of the 1$s$ state for $^{111}$Sn. We also find the shapes of the formation spectra by the ($d, ^3$He) reactions have the reasonable sensitivities to the $\sigma_{\pi N}$ value and we can expect to obtain extra information from the observed spectra by comparing them to the theoretical results.

Finally, we investigate the expected difficulties of the actual determination of the $\sigma_{\pi N}$ value and the correlation between potential parameters. We find there exists the correlation between $\sigma_{\pi N}$ and Re$B_0$ as in the case of the parameters $b_0$ and Re$B_0$~\cite{SM}. The correlation is expected to cause the difficulties to determine these parameters precisely in the analyses of the data of the deeply bound pionic atoms. We show in this article that the parameter correlations for the gap of the binding energies and the width indicate the different pattern in the Re$B_0 > 0$ region. The change of the neutron distribution of the nucleus is also found to affect the correlations. Hence, it would be necessary to perform the combined analyses of the various observables of the deeply bound pionic atoms for the different nuclei for which the neutron distributions are determined reliably.

\section*{ACKNOWLEDGEMENTS}
We appreciate the fruitful discussions with D.~Jido. The work was partly supported by JSPS KAKENHI Grant Numbers JP19K14709, JP16340083, JP18H01242, JP20KK0070, and by MEXT KAKENHI Grant Numbers JP22105517, JP24105712, JP15H00844.


\end{document}